# A Gravitational Lens Solution for IRAS F10214+4724

Tom Broadhurst
Department of Physics and Astronomy,
The Johns Hopkins University, Baltimore, MD 21218

Joseph Lehár
Department of Astronomy, Harvard University,
60 Garden St MS-78, Cambridge, MA 02138

## ABSTRACT

We show that the high redshift IRAS source F10214 is highly magnified by the gravitational field of an intervening elliptical galaxy, accounting for its many anomalous properties. Detailed radio and near-IR images identify the IRAS source with a symmetric arc, centered on a red object, or lensing galaxy. To explain the observed structures, the center of the source must much more highly magnified than its outer regions. Lensing predicts a small counterimage to the arc, which we find adjacent to the lensing galaxy. A red component in the observed spectrum suggests a lens redshift of unity, and the lens model yields a mass estimate of $M(r < 3\,\text{kpc}) \approx 10^{11}\,\text{M}_\odot$, consistent with an ordinary elliptical galaxy. We present new high-resolution optical images which show a thin arc of emission, implying an intrinsically small source ($< 0.5\,\text{kpc}$) which is highly magnified ($\sim 20\times$). Since the optical is strongly polarized with a Seyfert II spectrum, we propose that the optical arc is magnified image of the inner region of an obscured AGN. The obscuring "torus" will be similarly magnified, naturally accounting for the large IR flux. We show that finding objects like F10214+4724 in redshift surveys is probable, given the level of magnification bias expected for compact luminous IRAS sources. Such cases represent the obscured AGN counterparts to the lensed QSO population and, because of their extended sizes, are useful in determining the mass distribution in the lensing galaxies.

*Subject headings:* galaxies: individual (FSC10214+4724) — gravitational lensing

# 1. Introduction

The IRAS galaxy F10214 is unusual even by IRAS standards. It was discovered in a redshift survey of 1400 IRAS sources (Broadhurst *et al.* 1995) selected from the IRAS Faint Source Catalog (Moshir *et al.* 1989). At $z = 2.3$, F10214 has by far the largest redshift in the survey (Rowan-Robinson *et al.* 1991). The identification of the Far-IR flux with this redshift rests on the near coincidence between a mJy radio source and the high redshift optical object. Subsequently CO line emission at $z = 2.3$ was observed at this position for several transitions (Brown & Vanden Bout 1991, Solomon *et al.* 1992). This was taken to suggest a high molecular content, which, together with the large inferred IR luminosity ($\sim 3 \times 10^{13} \, h^{-2} \, L_\odot$, Downes *et al.* 1992), has led to much speculation about the nature of F10214. It is now clear that Seyfert II activity, not star formation, dominates the UV-optical emission. High excitation narrow lines and a strong degree of polarization in the rest UV have been found found (Lawrence *et al.* 1993, Elston *et al.* 1993). Lawrence *et al.* (1994) also report a non-detection of any significant X-ray emission and refute the hypothesis that F10214 is powered by a modestly extinguished QSO.

In this paper, we show how F10214 can be an ordinary obscured AGN which has been gravitationally lensed by a foreground elliptical galaxy. A characteristic of lensing is that the magnification depends strongly on the source size, and this allows us to understand the many unusual properties of F10214. We show in §2 that a simple ellipitical lens is all that is required to reproduce the observed morphology. We present new high resolution optical observations in §3, confirming the the lens interpretation. §4 describes what can be learned about the lensing galaxy, and in §5 we present a consistent explanation of the background source in terms of an obscured AGN. §6 addresses the *a priori* likelihood of finding systems like F10214 in the IRAS redshift survey, and finally in §7 we review our main conclusions. We will use the usual factor $h = H_0/100 \, \mathrm{km \, s^{-1} \, Mpc^{-1}}$ throughout, to account for the uncertainty in Hubble's constant.

# 2. F10214+4724 as a Gravitational Lens

In the lensing interpretation, the light from a distant IRAS source is magnified by the gravitational field of an intervening galaxy. Using the $K$-band images of Matthews *et al.* (1994), we identify the IRAS source with the arc-shaped object (Source 1), and the lensing galaxy with the central object (Source 2). The arc is focused on the lensing galaxy at a radius of $1\rlap{.}''2$, and contains a bright "core" which is dominated by line emission. Sources 3 and 4 cannot be additional lensed images of the IRAS source, and are probably just other galaxies close to the line of sight.

Only a very simple lens model is required to reproduce the observed $K$-band morphology. We use a singular isothermal lensing potential with an elliptical perturbation (Blandford & Kochanek 1987). The model parameters are: the lens center; a characteristic deflection angle or "ring radius" $\beta$; the ellipticity $\epsilon$; and the position angle $\phi$. We have placed a small source with a compact core just outside a cusp of the diamond caustic (see Figure 1). Two images of this source are formed: an arc-shaped image "A" to the South, and a smaller counterimage "B", just North of the lensing galaxy "G". The compact core gives rise to the bright point at the middle of the arc.

In our model, the magnification depends strongly on the source size, and this is what allows us to explain the unusual properties of F10214. Figure 2 shows how the magnification $m$ depends on the source size $r$. For small sources, $m(r)$ is dominated by the tangential stretching of A, so the arc length $\approx 2r\,m(r)$. The $K$-band arc is $\sim 2''$ across, suggesting a source with $r \sim 0\rlap{.}''2$ and $m \sim 5$. The 8 GHz radio source is coincident with the $K$-band core, and both are $\sim 0\rlap{.}''5$ across (Lawrence et al. 1993). This is much shorter than the $K$-band arc, and corresponds to a source with $r \sim 0\rlap{.}''01$ and $m \sim 50$. This large central magnification suggests that the Far-IR luminosity may also be magnified by a large amount. With $m \sim 50$, the bolometric luminosity falls within the range more typical of IRAS sources. Note that to explain the Far-IR luminosity, the central magnification is clearly preferred over the modest magnification of the larger $K$-band arc, since otherwise F10214 is made more unique, remaining the most luminous object in the Universe but *also* gravitationally lensed!

If F10214 is lensed, the lensing galaxy must be at a lower redshift than the source. Given the close projection of A and G, we have examined the published spectra of F10214, looking for evidence of the lensing galaxy. The spectra are too noisy for any absorption lines to be identified. However, a distinctive red continuum component is visible (Rowan-Robinson et al. 1991, Figure 2, Soifer et al. 1995, Figure 2) at wavelengths longer than $\sim 8000$ Å, similar to the spectrum of early-type galaxies at $z \sim 1$. We note that lensing galaxies are most likely to be early-type given their larger central densities (Turner et al. 1984).

Finally, the lens model predicts that there should be an opposing counterimage to the $K$-band arc, as illustrated in Figure 1. With this in mind, we have reanalyzed the $K$-band data of Matthews et al. 1994, kindly provided to us by the authors. Figure 3a shows the image, sharpened using the procedure described in §3. Both the arc "A" and the central galaxy "G" are present, but a tail of emission extends Northwards from G. Figure 3b shows the same map after removing G. A distinct source B is visible $\sim 0\rlap{.}''45$ North of G, providing a striking confirmation of the lensing interpretation.

## 3. New Optical Images of F10214+4724

We have imaged F10214 at high resolution in the optical B-band with the 4.2 m William Herschel Telescope in April 1994. Two ten-minute exposures, taken in $0''\!.6$ seeing, cover the source and two nearby bright stars. The images are highly oversampled, with $0''\!.1$ pixels, and the stars are both unsaturated and thus amenable to deconvolution. The number of iterations of the Lucy-Richardson algorithm was chosen to minimize the difference between the original and deconvolved images, after reconvolving with the stellar point spread function. Varying the number of iterations between 50 and 500, and using alternative template stars, produced no significant changes. Figure 4 shows the coadded deconvolved image of F10214.

A single bright arc is visible, coincident with the bright "core" of the $K$-band arc. Like the radio source, this arc is oriented E-W, with most of its flux toward the Western end. The arc is $\sim 1''$ long ($0''\!.5$ FWHM), so its inferred magnification from our model is $\sim 50$. The arc is unresolved in the N-S direction at $0''\!.2$ resolution, setting an upper limit on the source size of $r < 0.5\,h^{-1}$ kpc at $z = 2.3$. The magnitude of the optical arc is $B \approx 21.5$, and no counterimage is detected to a $3\sigma$ level of $B \sim 25$. This limits the optical magnification to $m > 20$, which is consistent with the arclength constraint. The lensing galaxy G is also not detected, so it must be very red ($B - K > 7.3$), consistent with an early-type galaxy at $z > 0.5$ (Mobasher *et al.* 1993).

## 4. Properties of the Lensing Galaxy

The near-IR geometry constrains the mass model of the lens. Fixing the lens position at G, the lensing mass must be elongated in order to produce both a long arc and a small counterimage. Since B is closer to G than A, the mass must be oriented with its major axis approximately N-S, and the arc symmetry forces $\phi \approx 0°$ (counter-clockwise from North). The positions of A, B, and G then constrain the ring radius to $\beta = 0''\!.91$, and the source position to $0''\!.38$ South of G. Finally, the ellipticity $\epsilon$ is determined by the requirement that the core of A be highly magnified. The size of this caustic increases with $\epsilon$, and extends to the source center when $\epsilon = 0.18$ (isophotal axial ratio $\sim 0.6$). After removing B, G shows no obvious elongation in the $K$-band image, but this is probably due to limited angular resolution.

The model can be used to estimate the lensing mass. For a lens at $z_L \sim 1$ and a source at $z_S = 2.3$, the ring size $\beta = 0''\!.91$ gives an isothermal velocity dispersion of $\approx 320\,\mathrm{km\,s^{-1}}$ (assuming a filled-beam cosmology with $\Omega = 1$). This corresponds to a mass of $\sim 3 \times 10^{11} \, h^{-1} \, M_\odot$ within $\beta$ ($\sim 4 \, h^{-1}$ kpc). This mass estimate is insensitive to assumptions about the mass model profile and the lens ellipticity, but varies by 30% for a lens redshift range of $0.8 < z_L < 1.2$.

This mass estimate can be compared with the luminosity of G. After removing the counterimage, the Near-IR magnitude is $K_s = 17.65$ (Matthews *et al.* 1994), and this converts to a standard $B_T$ magnitude if we assume an early-type galaxy spectrum. We use $B_T = K_s - \Delta K + (B_T - K) - E_{corr}$, where $B_T - K = 3.7$ is the present-date rest frame color of luminous early-type galaxies (Mobasher *et al.* 1993), and $\Delta K = 0.15(H - K) = 0.05$ is the correction between the standard $K$-band and the short $K_s$-band (James Graham, private communication). The K-corrected evolution in the redshifted $K$-band is small ($0.07 < E_{corr} < 0.12$ for early-type galaxies at $z \sim 1$, $0.5 < h < 1$, $\Omega = 1$; Bruzual & Charlot 1993). The resulting rest frame luminosity is then $M_{B_T} = -20.0 \pm 0.2$ for $0.5 < z_L < 1.5$, and the central mass-to-light ratio is $25.0 \pm 5.0 \, h$ in solar units. This is on the large side compared with local elliptical galaxies (Lauer 1985), but not surprising. A further check comes from the Faber-Jackson relation as expressed by Fukugita & Turner (1991). The expected velocity dispersion, $\sigma \approx 255 \, h^{0.5} \, \mathrm{km \, s^{-1}}$, is consistent with the lens dispersion if a factor of $\sqrt{1.5}$ is included to account for the stellar brightness profile (Gott 1977).

## 5. The Nature of the Background Source

Gravitational lensing can explain the morphology and unusual flux levels in F10214, but it places severe constraints on the nature of the source. To achieve the high magnifications in our model, the source radius $r$ cannot exceed the caustic offset $b$, and since $\mu(b) \sim 2\beta/b$ (see Figure 2), $r < 2\beta/m$. There is a lower limit to the source size if the Far-IR source is thermal. About half of the bolometric flux can be fitted by an 80 K thermal component of the rest frame IR spectrum (Downes *et al.* 1992). In this case, the apparent luminosity is magnified, so that $L \propto mR^2T^4$. For F10214, the minimum blackbody radius is $R_{BB} \approx 560 \, m^{-0.5} \, h^{-1}$ pc, corresponding to $r > 0\rlap{.}''14 \, m^{-0.5}$ at $z = 2.3$. So the allowed range is $0\rlap{.}''14 \, m^{-0.5} < r < 1.8 \, m^{-1}$, which limits the magnification to $m < 170$. For $m = 50$, the range is $0\rlap{.}''02 < r < 0\rlap{.}''04$, corresponding to a source diameter of $\sim 100 \, h^{-1}$ pc at $z = 2.3$. Thus if the lens magnification can reduce the intrinsic luminosity of F10214 to within the observed IRAS range, the IR source must be very small, and optically thick if it is thermal.

The key to understanding F10214 is its AGN nature. The Optical/IR spectrum is clearly Seyfert II like, and is highly polarized ($\approx 15\%$) (Lawrence *et al.* 1993; Elston *et al.* 1993). Such high levels of polarization are found in local Seyferts, in the central "mirror" region (Antonucci 1993). The size of local AGN mirrors is $\approx 50 - 100$ pc. This is consistent with the above limits, and for $m \sim 50$, would produce a $\sim 1''$ long optical arc. Since the optical flux is highly magnified (see §3), and consistent with the size of local AGN mirrors, we conclude that the optical arc is a highly magnified image of the central region of an obscured AGN.

If F10214 is a lensed AGN, then we may also expect to see highly magnified IR emission from the dusty "torus" obscuring the AGN. The torus would radiate predominantly in the IR, peaking at around $\sim 10 - 20\mu$, on a size scale of $10 - 100\,\mathrm{pc}$ (Pier & Krolik 1993). This region is small enough and optically thick, satisfying the blackbody constraints. Furthermore, such torus models can be constructed to agree well with the overall IR spectra of luminous IRAS galaxies for only modest dust masses $M_{dust} \sim 10^{4-5} M_\odot$, and also reproduce well the spectrum of F10214 (e.g. Figure 6 of Granato & Danese 1994). If the AGN in F10214 is similar to the local Seyfert II NGC1028, with a total IR luminosity of $\sim 5 \times 10^{11}\,\mathrm{L}_\odot$ (Rieke and Low 1975), then $m \sim 25 - 50$ brings this close to the apparent (i.e lensed) bolometric luminosity of F10214.

The view that F10214 is a lensed AGN solves several other problems associated with this source. (1) High magnification can explain the anomalously large CO emission. Although the temperature deduced from the CO transitions is abnormally high for giant molecular clouds (Solomon *et al.* 1995), it is consistent with the conditions in the central region of AGNs (Meixner *et al.* 1990). In this case, the CO velocity profile reflects the dynamics of the central region. (2) If the AGN narrow line region has any ionization structure, the spectral lines will be magnified by different amounts. This effect may explain the anomalously large high ionization lines like NV in F10214, compared to other Seyfert II's. (3) The unexpected polarization angle is also explained. The **E** vector is usually perpendicular to the major radio axis in Seyfert II sources. However, in the case of F10214 the polarization is only 20° from the major axis of the radio emission (Lawrence *et al.* 1993). In our model the radio source elongation is dominated by lensing, so the background radio source could be in any intrinsic orientation.

## 6. Probability of Finding Sources Like F10214+4724

One might ask how likely is such a lensing event, given what is known about the IRAS luminosities and the intervening galaxy population. This is easy to quantify because the IRAS redshift survey in which F10214 was found has a pure flux limit of $0.2\,\mathrm{Jy}$ at $60\mu$, and angular coverage of $0.20\,\mathrm{sr}$ (Broadhurst *et al.* 1995). Note that gravitational lensing introduces a significant "magnification bias" in a Far-IR flux-limited survey, favoring highly magnified sources like F10214.

We determine an optical depth to lensing in the limit of compact sources following Turner et al. (1984), and include the extra boost to the high magnification tail from the diamond caustics of elliptical lenses (Wallington & Narayan 1993). The optical depth is sensitive to the choice of IRAS galaxy evolution, the form of which is constrained by the 0.2 Jy redshift survey from which F10214 was selected, albeit at lower redshift $z < 0.4$ (Broadhurst et al. 1995). Pure density evolution at a rate of $(1+z)^5$ is preferred, in which case we expect 0.3 lensed IRAS sources over the 0.20 sr of the redshift survey and the expected source redshift is $<z_S> \approx 1.2$. Pure luminosity evolution at a rate of $(1+z)^{2.5}$ is also a good fit to the evolution, yielding 1.2 expected lensed sources, and $<z_S> \approx 2.0$. These probabilities are encouragingly large, and it seems likely that more systems like F10214 will be found.

This calculation also produces expectation values for the source magnification and the lens redshift. The latter depends only on the source redshift, and we find $<z_L> = 0.7 \pm 0.2$ with the dispersion derived from the probability distribution. Source magnifications of $\sim 25$ and $\sim 40$ are expected for pure luminosity and density evolution respectively, in the limit that luminous IRAS sources are small.

## 7. Conclusions

Gravitational lensing clearly provides a natural explanation for all of the unusual properties of the IRAS source F10214+4724. We have constructed a simple lens model to reproduce the observed morphology, and have shown that the model is consistent with an ordinary early-type galaxy. We presented new optical images, which show that core of F10214 must be magnified by $m > 20$. The high core magnification and its spectral and polarization properties led us to conclude that the source is an obscured AGN. We explained the large IR flux as thermal emission from the obscuring torus, magnified $\sim 50\times$, which places the intrinsic flux of F10214 inside the observed range of luminous IRAS galaxies. Finally, we showed that the a priori probability of finding cases like F10214 is encouragingly large, given the level of magnification bias for compact Far-IR selected sources.

We believe that systems like F10214 represent those lensed AGNs whose central engines are obscured. If they are as common as we predict, lensed AGNs rather than lensed QSOs could be more useful in probing the lens galaxy mass distributions, because the background emission is measurably extended. The IRAS source with the next highest inferred luminosity is F15307+3252, at $z = 0.93$ (Cutri et al. 1994), and it is probably another case like F10214. The lensed QSO 2237+030 ($z_S = 2.5$) is a strong CO emitter (Barvainis et al. 1995), and may be a face-on version of F10214.

Current searches for lensing have used optical QSOs and active radio sources, which are only a subset of the full AGN population. Selection by Far-IR allows a fuller coverage of AGNs but is presently limited by the relatively low IRAS flux limit. In the future, lensed AGN's may be efficiently selected by correlating hard X-ray, Radio, and Far-IR sky surveys with the multicolor Sloan Digital Sky Survey, selecting out hybrid red+blue identifications for detailed HST follow-up.

More work is clearly needed on F10214 itself. Deep HST images should confirm the counterimage, and place tighter limits on the source size from the radial width of the main arc. Ring structures should also be detectable at low isophotes, corresponding to the outer galaxy of the source, constraining the lensing mass profile at larger radii. In addition, the light profile of the lensing galaxy can be determined in more detail, to be compared with the mass profile. Since they are highly magnified, any structures in the arc could give very detailed information on the background galaxy. At other frequencies, it should also be possible to detect more directly the AGN in reflected broad absorption lines and from the unabsorbed hard X-ray emission. Given improvements in high frequency interferometers, high resolution CO maps might provide useful constraints on the lensing mass distribution, since the CO emission is extended. If the lensing galaxy is a luminous early-type galaxy at $z_L \sim 1$ as we suggest, obtaining its redshift should be relatively straightforward. The lens redshift would allow us to improve the mass estimate, and would provide conclusive confirmation of lensing.

The authors would like to thank Tim Heckman and Julian Krolik and Ed Turner for useful suggestions and Dave Carter and Karl Glazebrook for help with the WHT observations.

---

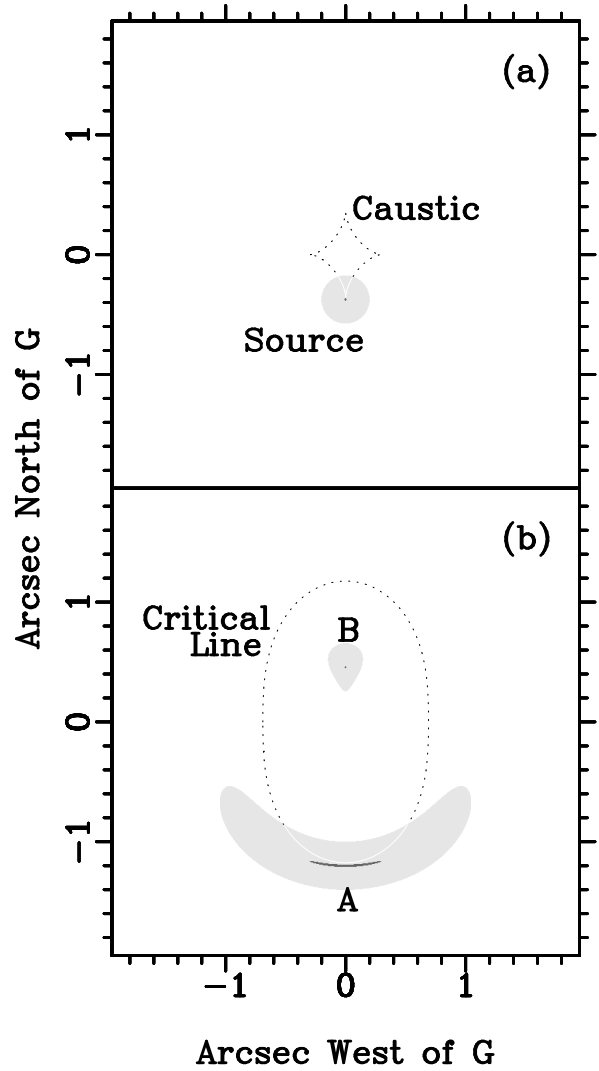

**Figure 1:** Gravitational lens model for F10214+4724. (*a*) shows how the source would appear in the absence of the lens, and (*b*) shows the distorted images seen through the lens. Sources outside the caustic produce two images, and those inside produce four. The lens model (Blandford & Kochanek 1987) is centered on G, with $\beta = 0\farcs91$, $\epsilon = 0.18$, and $\phi = 0°$ (CCW from North). The small source ($r = 0\farcs2$) has a compact core ($r = 0\farcs01$), which is offset by $b = 0\farcs024$ from a cusp in the caustic.

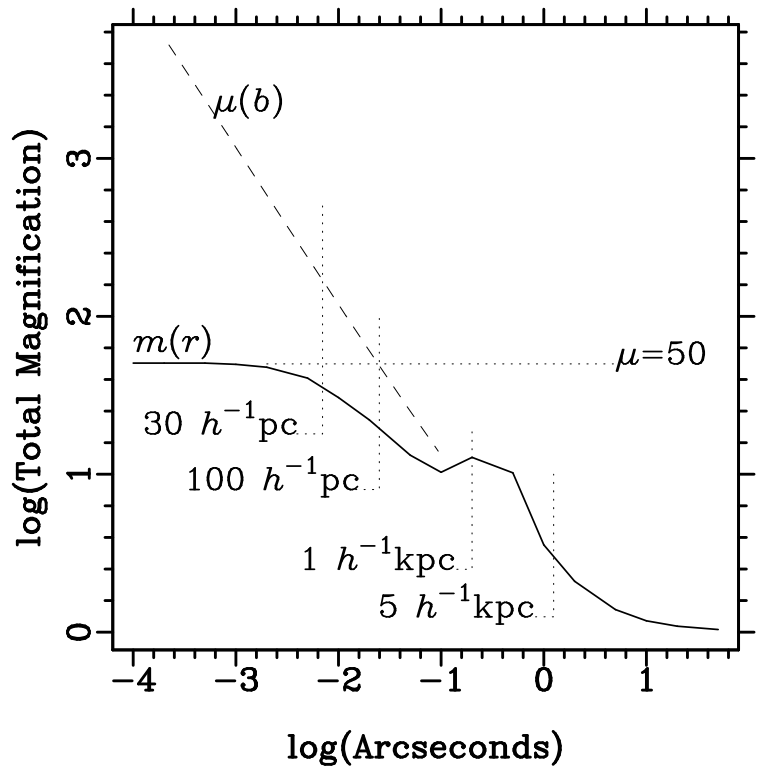

**Figure 2:** Total magnification $m(r)$ as a function of source size, empirically determined for the model in Figure 1. For a caustic offset $b = 0\rlap{.}''024$, $m(r)$ increases as $r$ decreases, reaching a maximum $\mu = 50$ when $r < b$. The empirically determined $\mu(b)$ is also shown. The bump at $r \sim 0\rlap{.}''3$ is due to the appearance of ring-like images when the source touches all four folds of the caustic. Source sizes are shown, assuming $z = 2.3$ ($1'' \sim 4\,\mathrm{kpc}$).

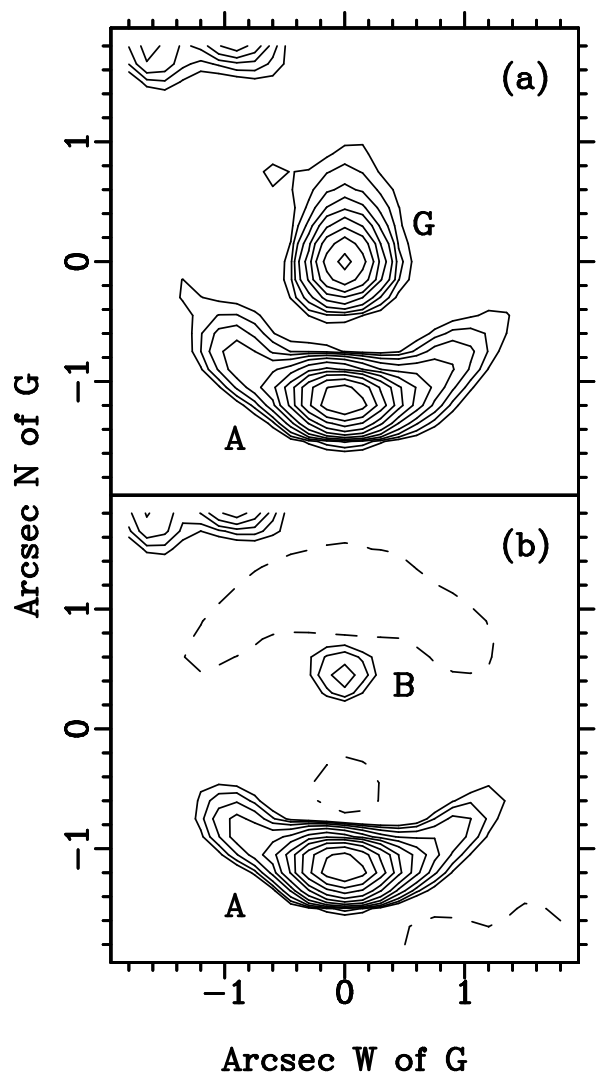

**Figure 3:** $K$-band image of F10214+4724, from the data of Matthews *et al.* (1994). (*a*) shows the image sharpened to $0\rlap{.}''3$ resolution, showing the arc "A" and the central galaxy "G". (*b*) is the same image rotated by 180° around G, and subtracted from (*a*), to remove symmetric emission from G. The arc counterimage "B" is visible $\sim 0\rlap{.}''45$ North of G. The pixel size is $0\rlap{.}''15$, and the contours mark fractional intervals of $\sqrt{2}$ in intensity, starting at the map maximum.

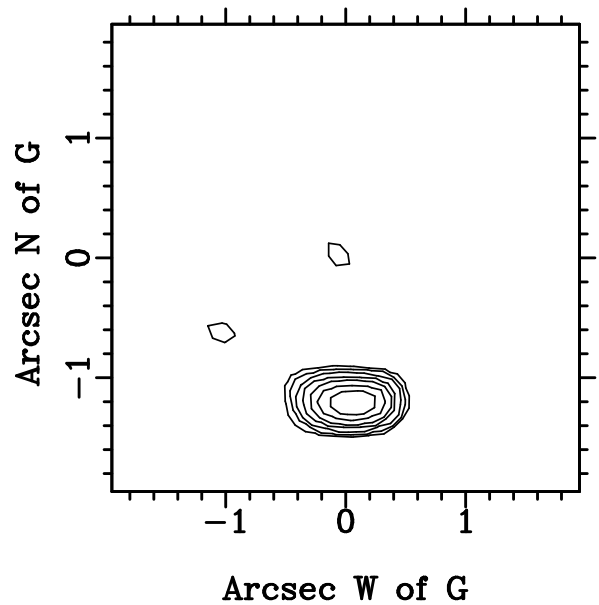

**Figure 4:** *B*-band image of F10214+4724, sharpened to 0″.2 resolution. The source is coincident with A in Figure 3, and is unresolved in the N-S direction. The pixel size is 0″.1, and the seeing was $\sim 0″.6$. The source magnitude is $B \approx 21.5$, with a sky level of $\approx 21.5\,\mathrm{mag\,arcsec^{-2}}$. An unresolved source with $B \sim 25$ would give a $3\sigma$ detection. The contours mark fractional intervals of $\sqrt{2}$ in intensity, starting at the map maximum.